\documentclass[conference]{IEEEtran}
\IEEEoverridecommandlockouts

\usepackage{cite} 
\usepackage{amsmath,amssymb,amsfonts}
\usepackage{algorithmic}
\usepackage{graphicx} 
\usepackage{textcomp}
\usepackage{xcolor}
\usepackage{subcaption}
\usepackage{bm} 
\usepackage[bookmarks=false]{hyperref} 
\usepackage{multirow} 

\usepackage{booktabs} 
\usepackage{placeins} 
\usepackage{balance} 
\usepackage{footnote} 
\usepackage{tabularx} 
\usepackage{stfloats} 
\usepackage{url}
\hypersetup{
    hidelinks,        
    breaklinks=true   
}

\newcommand{\C}{\mathbb{C}}
\newcommand{\R}{\mathbb{R}}
\newcommand{\vect}[1]{\mathbf{#1}} 
\newcommand{\mat}[1]{\mathbf{#1}} 
\DeclareMathOperator*{\argmax}{arg\,max} 
\def\j{\mathrm{j}}

\def\BibTeX{{\rm B\kern-.05em{\sc i\kern-.025em b}\kern-.08em
    T\kern-.1667em\lower.7ex\hbox{E}\kern-.125emX}}

\newcommand{\degree}{\ensuremath{^\circ}}

\begin{document}
\title{{CFARNet: Learning-Based High-Resolution Multi-Target Detection for Rainbow Beam Radar}
\thanks{This work is supported by the National Natural Science Foundation of China (NSFC) under Grant 62125108 and Grant 62431014, and in part by the Science and Technology Commission Foundation of Shanghai under Grant 24DP1500702. The source code of CFARNet is publicly available at \url{https://github.com/SJTU-WirelessAI-Lab/CFARNet}.}}

\author{
    \IEEEauthorblockN{Qiushi Liang, Yeyue Cai, Jianhua Mo, and Meixia Tao}
    \IEEEauthorblockA{Dept. of Electronic Engineering, Shanghai Jiao Tong University, Shanghai, China\\
    Emails: \{lqs020206, caiyeyue, mjh, mxtao\}@sjtu.edu.cn}
}

\maketitle
\IEEEpeerreviewmaketitle 

\begin{abstract}
Millimeter-wave (mmWave) OFDM radar equipped with rainbow beamforming, enabled by phase-time arrays (PTAs), provides wide-angle coverage and is well-suited for fast real-time target detection and tracking. However, accurate detection of multiple closely spaced targets remains a key challenge for conventional signal processing pipelines, particularly those relying on constant false alarm rate (CFAR) detectors.
This paper presents CFARNet, a learning-based processing framework that replaces CFAR with a convolutional neural network (CNN) for peak detection in the angle-Doppler domain. The network predicts target subcarrier indices, which guide angle estimation via a known frequency-angle mapping and enable high-resolution range and velocity estimation using the MUSIC algorithm.
Extensive simulations demonstrate that CFARNet significantly outperforms a baseline combining CFAR and MUSIC, especially under low transmit power and dense multi-target conditions. The proposed method offers superior angular resolution, enhanced robustness in low-SNR scenarios, and improved computational efficiency, highlighting the potential of data-driven approaches for high-resolution mmWave radar sensing.
\end{abstract}


\section{Introduction}

Integrated sensing and communication (ISAC) has emerged as a pivotal paradigm for 6G systems, aiming to unify sensing and communication functionalities by sharing hardware and spectral resources to improve efficiency and enable novel capabilities \cite{Gonzalez_VTM25, Jiang_Yihang_MCOM25}. Employing communication waveforms such as orthogonal frequency division multiplexing (OFDM) for radar sensing is a key enabler for ISAC, promoting spectral coexistence and hardware reuse \cite{Liu_Fan_JSAC22}.

Phase-time arrays (PTAs) \cite{ratnam2022joint, Alammouri_Access22, Yildiz_VTC24, Mo_Jianhua_Asilomar24, Cai_Yeyue_WCNC25, cai2025hybrid, Nam2025Joint} represent a new millimeter-wave (mmWave) MIMO architecture that combines phase shifters (PSs) and true-time delay (TTD) elements \cite{rotman2016true, Forbes_JSSC23}. PTAs facilitate frequency-dependent beamforming strategies such as rainbow beams or controllable beam squint beamforming (CBS-beamforming) \cite{Yan_Han_Asilomar19, zhai2021ss, Cui_Mingyao_TWC23, Wadaskar_TWC25, gao2023integrated, Luo2024yolo}, where different OFDM subcarriers are deliberately steered toward distinct angles across a wide sector. This approach enables rapid angular scanning using a single OFDM symbol \cite{gao2023integrated, Luo2024yolo, Zhou_Gui_WCNC24, Zhou_Gui_JSAC25, Lei2025Deep, Zheng2025Near}, drastically reducing the sensing overhead compared to traditional beam sweeping methods that sequentially scan directions \cite{Mo2019Beam, Heng2021Six}. The beam squint effect, usually viewed as a drawback in wideband systems, is instead exploited here to encode angular information into the frequency domain.

While rainbow beamforming improves sensing efficiency, accurately estimating multiple mobile targets’ kinematic parameters (angle, range, and radial velocity) from received echoes remains challenging. Recent radar signal processing methods of PTAs \cite{Luo2024yolo, Zhou_Gui_WCNC24, Zhou_Gui_JSAC25} typically identify target angles based on the subcarrier with the maximum echo power, followed by super-resolution techniques such as MUSIC \cite{MUSIC_Schmidt_Ref} or ESPRIT \cite{ESPRIT_Roy_Ref} to estimate range and radial velocity. Specifically, the YOLO framework in \cite{Luo2024yolo} follows this pipeline: it first transforms the received signal into the angle-Doppler domain via FFT, detects targets using constant false alarm rate (CFAR) \cite{CFAR_Overview_Ref} detection, and subsequently applies MUSIC for fine-grained range and velocity estimation.

However, the CFAR stage often constitutes the performance bottleneck in conventional radar signal processing pipelines. CFAR relies on estimating local clutter statistics to set adaptive detection thresholds, which becomes problematic when targets are closely spaced in angle and Doppler, as their responses may merge or interfere with neighboring cells \cite{CFAR_Overview_Ref}. Furthermore, CFAR is especially vulnerable when targets have significantly different reflection strengths and tends to perform poorly under non-homogeneous or rapidly varying clutter environments \cite{CFAR_Limitations_Ref}. These limitations can lead to missed detections, merged targets, or inaccurate peak localization, thereby degrading the overall parameter estimation accuracy.

In this work, we propose \textbf{CFARNet}, a convolutional neural network (CNN)-based framework for enhanced multi-target parameter estimation in mmWave rainbow beam OFDM radar systems. CFARNet replaces the conventional CFAR detector with a trained CNN that directly predicts the subcarrier indices corresponding to target peaks from the noisy angle-Doppler spectrum. These predicted indices guide the subsequent parameter estimation: angles are derived from the known frequency-angle mapping of the rainbow beam, while range and radial velocity are estimated via the MUSIC algorithm applied to an angle-Doppler map centered on the predicted angles. We conduct a comprehensive comparison against the CFAR-based baseline across varying numbers of targets, angular separations, and transmit power levels. Simulation results demonstrate that CFARNet substantially improves estimation accuracy, particularly under low transmit power and in dense multi-target scenarios.


\section{System Model}
\label{sec:SystemModel}
\begin{figure}[t]
    \centering
    \includegraphics[width=\linewidth]{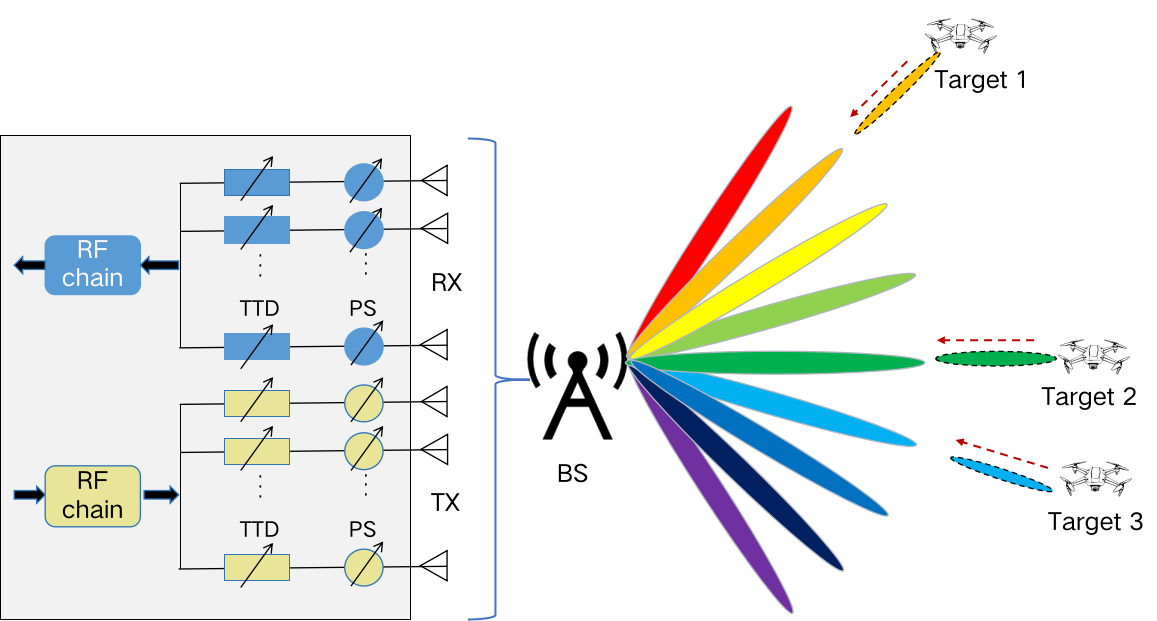} %
    \caption{System architecture of phase-time array rainbow beamforming for multi-target sensing.}
    \label{fig:system}
    \vspace{-3mm}
\end{figure}

We consider a mmWave base station (BS) performing target sensing, equipped with colocated transmitter and receiver units, as illustrated in Fig.~\ref{fig:system}. Each unit employs a single RF chain connected to a uniform linear array (ULA) comprising $N$ antenna elements with half-wavelength spacing, i.e., $d = \lambda_c/2 = c/(2f_c)$, where $\lambda_c$ is the wavelength corresponding to the carrier frequency $f_c$. 

The system adopts an OFDM waveform with $M=2048$ subcarriers, indexed by $m \in \{0, 1, \dots, M-1\}$. The total bandwidth is $W$, resulting in a subcarrier spacing of $\Delta f = W/M$. The frequency of the $m$-th subcarrier is given by $f_m = f_0 + m\Delta f$, where $f_0 = f_c - W/2$ denotes the frequency of the first subcarrier. The duration of each OFDM symbol is $T_s = 1/\Delta f$. During sensing, the BS transmits and receives $N_s$ consecutive OFDM symbols.

\subsection{Rainbow Beamforming}
A rainbow beamforming strategy is employed to enable simultaneous sensing of multiple angles. This strategy leverages pre-determined PS and TTD settings applied to each antenna element $n$, aiming to create a frequency-dependent beam pattern that maps subcarriers across the bandwidth $W$ to distinct angles within a predefined sector $[\phi_{\text{start}}, \phi_{\text{end}}]$.

Let $\tilde{f}_m = m \Delta f = m W/M$ denote the baseband frequency corresponding to the $m$-th subcarrier. The effective beamforming weight vector $\vect{w}_m \in \C^{N\times 1}$ for the $m$-th subcarrier, which remains constant across the $N_s$ OFDM symbols, is given by
\begin{equation}\label{eq:BeamformingWeightDef_Ref}
[\vect{w}_m]_n = \frac{1}{\sqrt{N}} \exp(-\j 2\pi \phi_n) \exp(-\j 2\pi \tilde{f}_m t_n),
\end{equation}
where $\phi_n$ and $t_n$ represent the phase shift and time delay applied at the $n$-th antenna element, respectively.

To steer the beam from $\phi_{\text{start}}$ at the starting frequency $f_0$ to $\phi_{\text{end}}$ at the highest frequency $f_M = f_0 + W$, the PS and TTD settings are designed as follows \cite[Sec. III.B]{Luo2024yolo}:
\begin{align}
\phi_n &= -\frac{f_0 n d \sin\phi_{\text{start}}}{c}, \label{eq:PS_setting_Ref} \\
t_n &= \frac{n d}{W c} \left( f_0 \sin\phi_{\text{start}} -\left(f_0 + W \right) \sin\phi_{\text{end}} \right). \label{eq:TTD_setting_Ref_expanded}
\end{align}



\subsection{Received Echo Signal}
We consider a scenario with $K$ distinct mobile targets located in the far field. The $k$-th target is characterized by its angle $\phi_k$ relative to the ULA broadside, its initial range $r_k$ from the array center, and its constant radial velocity $v_{r,k}$, which is positive for receding targets. Each target is associated with a reflectivity coefficient $\alpha_k$, which accounts for factors such as the Radar Cross Section (RCS) and round-trip propagation path loss. The reflectivity $\alpha_k$ is modeled as
\begin{equation}\label{eq:alpha_k_calc_rcs}
\alpha_k = \sqrt{\frac{\lambda_c^2 G_T(\phi_k) G_R(\phi_k) \sigma_{c,k}}{(4\pi)^3 r_k^4}},
\end{equation}
where $G_T(\phi_k)$ and $G_R(\phi_k)$ denote the element gains of the transmit and receive antennas at the BS, respectively, $r_k$ is the target range, and $\sigma_{c,k}$ is the RCS of the $k$-th target. 

The round-trip channel matrix $\mat{H}_{n_s,m} \in \C^{N \times N}$, capturing the propagation effects between the transmit and receive arrays for subcarrier $m$ at symbol index $n_s$, is expressed as
\begin{equation}\label{eq:OverallChannel}
\mat{H}_{n_s,m} = \sum_{k=1}^K \alpha_k e^{-\j 2\pi f_m \frac{2 r_k}{c}} e^{\j \frac{4\pi v_{r,k} f_0}{c} n_s T_s}\vect{a}_m(\phi_k) \vect{a}_m^H(\phi_k),
\end{equation}
where the round-trip propagation delay and Doppler shift effects are incorporated, and $\vect{a}_m(\phi_k) \in \C^{N\times1}$ is the array steering vector corresponding to angle $\phi_k$ at frequency $f_m$, defined as
\begin{align}
[\vect{a}_m(\phi_k)]_n &= \exp\left(\j 2\pi f_m \frac{n d \sin \phi_k}{c}\right), \label{eq:ArrayVector} 
\end{align}

The received baseband signal sample $y_{n_s,m}$ at the output of the receive beamformer for subcarrier $m$ and symbol $n_s$ is
\begin{equation}\label{eq:ReceivedSignalMono}
y_{n_s,m} = \sqrt{P_t}  \vect{w}_m^H \mat{H}_{n_s,m} \vect{w}_m x_{n_s,m} + \eta_{n_s,m},
\end{equation}
where $P_t$ denotes the transmit power, $\eta_{n_s,m}$ represents the complex additive white Gaussian noise (AWGN), and $x_{n_s,m}$ is the known transmitted pilot symbol. For simplicity, we assume $x_{n_s,m} = 1$ for all $(n_s, m)$.
The collection of received samples forms the data matrix $\mat{Y}_{\text{echo}} \in \C^{N_s \times M}$. We assume that static clutter filtering has been applied, and thus $\mat{Y}_{\text{echo}}$ contains only the echo signals from mobile targets.

\section{YOLO Baseline: CFAR + MUSIC}
Before presenting the proposed method, we first introduce the YOLO baseline, which follows a radar signal processing pipeline conceptually similar to \cite{Luo2024yolo}. The processing steps are summarized as follows:
\begin{enumerate}
    \item \textbf{Preprocessing:}
    The angle-Doppler spectrum $\mat{Y}_{\text{AD}}$ is computed by applying an $N_s$-point discrete Fourier transform (DFT) along the temporal dimension of $\mat{Y}_{\text{echo}}$:
    \begin{equation}
    \mat{Y}_{\text{AD}} = \mat{D}_{N_s} \mat{Y}_{\text{echo}},
    \end{equation}
    where $\mat{D}_{N_s}$ denotes the $N_s$-point DFT matrix.
    
    \item \textbf{Peak Detection by CFAR:}  
    The magnitude of $\mat{Y}_{\text{AD}}$ is first obtained, denoted as $\mat{G}_{\text{AD}}$, where $[\mat{G}_{\text{AD}}]_{n_s, m} = \left|\left[\mat{Y}_{\text{AD}}\right]_{n_s, m}\right|$.  
    A two-dimensional cell-averaging CFAR (CA-CFAR) detector \cite{CFAR_Overview_Ref} is then applied over $\mat{G}_{\text{AD}}$.  
    Specifically, for each element under test, its power is compared against an adaptive threshold $\epsilon P_n$, where $P_n$ represents the estimated local noise power averaged over surrounding reference cells excluding guard cells, and $\epsilon$ is a scaling factor determined by the desired probability of false alarm $P_\text{FA}$. An element is declared a peak if its power exceeds the threshold.
    
    \item \textbf{Target Estimation:}  
    Let $m_{\text{CFAR},k}$ denote the subcarrier index corresponding to the $k$-th detected peak. The angle $\hat{\phi}_k$ is calculated using Eq.~(\ref{eq:angle_formula}), and the range $\hat{r}_k$ and radial velocity $\hat{v}_{r,k}$ are estimated using the MUSIC algorithm, which will be detailed in Sec.~\ref{subsec:MUSIC}.
\end{enumerate}

While the YOLO baseline performs well with well-separated targets, its accuracy degrades significantly when targets are closely spaced in angle and Doppler domains. In these cases, CFAR often fails to distinguish nearby targets due to overlapping responses and limited resolution, motivating the development of a learning-based approach for more robust multi-target detection.

\section{Proposed Pipeline: CFARNet + MUSIC}
The proposed pipeline replaces the CFAR detection stage in the baseline with a trained convolutional neural network (CNN), referred to as CFARNet. The detailed processing steps are as follows:

\begin{figure}[t]
    \centering
    \includegraphics[width=1\linewidth]{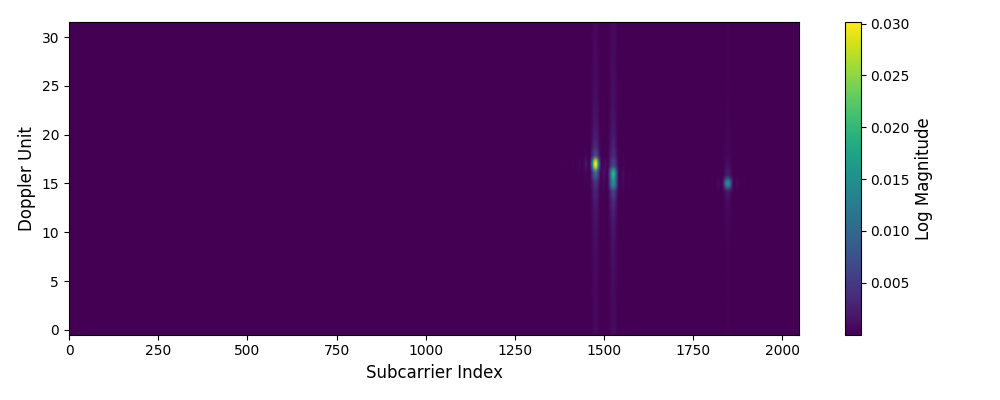} 
    \caption{An example of the angle-Doppler log-magnitude spectrum $\mat{L}_{\text{AD}}$. There are 3 targets, $M=2048$ subcarriers and $32$ Doppler indices.}
    \label{fig:input_heatmap}
    \vspace{-3mm}
\end{figure}

\begin{enumerate}
    \item \textbf{Preprocessing:}  
    The angle-Doppler spectrum $\mat{Y}_{\text{AD}}$ is first computed from the received echo matrix $\mat{Y}_{\text{echo}}$, following the same procedure as in the baseline.  
    Subsequently, the log-magnitude of the angle-Doppler spectrum is computed as the input to the CFARNet network:
    \begin{align}
        \left[\mat{L}_{\text{AD}} \right]_{n_s, m} &= \log\left(1 + \left[\mat{G}_{\text{AD}}\right]_{n_s, m}\right) \\
        &= \log\left(1 + \left|\left[\mat{Y}_{\text{AD}} \right]_{n_s, m}\right|\right).
    \end{align}
    An example of $\mat{L}_{\text{AD}}$ is shown in Fig.~\ref{fig:input_heatmap}.  
    The logarithmic transformation compresses the wide dynamic range inherent in radar signals by enhancing the visibility of weaker targets relative to stronger targets or noise, and stabilizes the training process of the neural network.
    
    \item \textbf{Peak Prediction by CFARNet:}
    The CFARNet model takes the log-magnitude angle-Doppler spectrum $\mat{L}_{\text{AD}}$ as input and outputs a logit vector $\vect{z} \in \R^{M}$ indicating the likelihood of a peak at each subcarrier.
    
    \item \textbf{Top-K Peak Selection:}  
    Select the $K$ subcarrier indices ${m_{\text{CNN}, k}}$ with the highest values in the logit vector $\vect{z}$ as the predicted peak locations.
    
    \item \textbf{Target Estimation:}  
    Estimate angle, range, and velocity from the predicted subcarrier indices using the same method as the YOLO baseline's Step 3, which will be detailed in Sec.~\ref{subsec:MUSIC}.
\end{enumerate}

\subsection{CFARNet Architecture}
\label{subsubsec:CNNArch}
\begin{figure}[t]
    \centering    \includegraphics[width=0.7\linewidth]{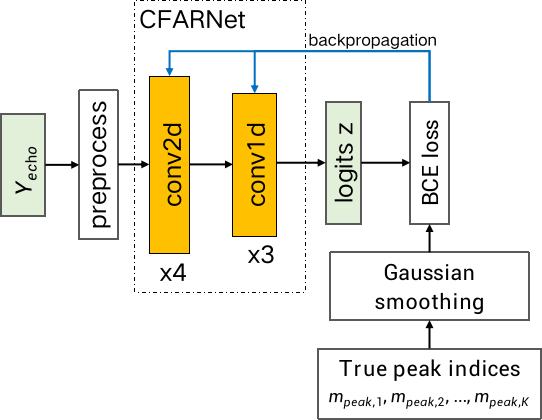} 
    \caption{Architecture of the proposed CFARNet network. }
    \label{fig:cnn_arch}
    \vspace{-3mm}
\end{figure}

The CFARNet network architecture, illustrated in Fig. \ref{fig:cnn_arch}, processes the log-magnitude angle-Doppler spectrum. Key layers include:
\begin{itemize}
    \item \textbf{Feature Extractor:} Four 2D convolutional blocks (Conv2D, BatchNorm, LeakyReLU, Dropout) designed to extract features across Doppler and subcarrier dimensions. The blocks progressively reduce the Doppler dimension while maintaining the subcarrier dimension $M$, increasing channel depth: $1 \rightarrow 64 \rightarrow 128 \rightarrow 256 \rightarrow 512$.
    \item \textbf{Pooling:} Max pooling along the reduced Doppler dimension to aggregate features.
    \item \textbf{Predictor Head:} Two 1D convolutional blocks (Conv1D, BatchNorm, LeakyReLU, Dropout) operating along the subcarrier dimension, followed by a final Conv1D layer that outputs $M$ logits, one for each subcarrier index.
\end{itemize}

The CFARNet network is trained as a multi-label classification problem across the $M$ subcarrier indices. We use the binary cross-entropy (BCE) with Logits loss, which combines a sigmoid activation with the BCE loss for numerical stability. Crucially, instead of using sparse one-hot encoded vectors representing the ground truth peak indices $\{m_{\text{peak}, k}\}$, we employ Gaussian label smoothing. This creates a soft target vector $\vect{y}_{\text{target}} \in \R^{M}$, providing richer gradient information around the true peaks and improving robustness to slight localization ambiguities.

For a given sample with valid ground truth peak indices $\mathcal{M}_v = \{m_{\text{peak}, 1}, \dots, m_{\text{peak}, K}\}$, the smoothed target value for subcarrier index $m$ is generated by summing Gaussian functions centered at each true peak and clamping the result between 0 and 1:
\begin{equation}\label{eq:gaussian_target}
y_{\text{target}, m} = \min\left(1, \sum_{k \in \mathcal{M}_v} \exp\left( -\frac{(m - m_{\text{peak}, k})^2}{2 \sigma_{\text{loss}}^2} \right) \right),
\end{equation}
where $\sigma_{\text{loss}}$ controls the width of the Gaussian smoothing.

Let $\vect{z} = [z_0, \dots, z_{M-1}]$ be the raw logit output vector from the CFARNet predictor head for a given input sample. The BCE loss is effectively calculated as:
\begin{multline}\label{eq:bce_loss}
\mathcal{L}_{\text{BCE}} = -\frac{1}{M} \sum_{m=0}^{M-1} \Big[ y_{\text{target}, m} \log(\sigma(z_m)) \\
 + (1 - y_{\text{target}, m}) \log(1 - \sigma(z_m)) \Big],
\end{multline}
where $\sigma(z_m) = 1 / (1 + \exp(-z_m))$ is the sigmoid function. This setup encourages the network to output high logits at and near the true peak locations defined by the smoothed target $\vect{y}_{\text{target}}$.

The CFARNet network was trained using the AdamW optimizer with learning rate $10^{-4}$ and weight decay rate $10^{-5}$, a CosineAnnealingLR scheduler, and early stopping based on validation loss to prevent overfitting. 

\subsection{Angle, Range, Radial Velocity Estimation} \label{subsec:MUSIC}
\subsubsection{Angle Estimation}
\label{subsubsec:AngleFormula}
The estimated angle $\hat{\phi}_k$ corresponding to a detected peak subcarrier index $m_p \in \{0, \dots, M-1\}$ is determined by the known frequency-angle mapping of the rainbow beamforming setup. Following \cite[Eq. (12)]{Luo2024yolo}, this relationship can be expressed as:
\begin{equation}\label{eq:angle_formula}
    \sin \hat{\phi}_{k} = \frac{(W-\tilde{f}_{m_p})f_{0}}{Wf_{m_p}}\sin \phi_{\mathrm{start}}+\frac{(W+f_{0})\tilde{f}_{m_p}}{Wf_{m_p}}\sin \phi_{\mathrm{end}},
\end{equation}
where $f_0 = f_c - W/2$, $\tilde{f}_{m_p} = m_p \Delta f$, $f_{m_p} = f_0 + \tilde{f}_{m_p}$.

\subsubsection{MUSIC Algorithm for range and velocity estimation}
The MUSIC algorithm is employed for high-resolution range and velocity estimation using the signal sub-band $\mat{Y}_{\text{sub}} = \mat{Y}_{\text{AD}}[:, m_{p}-W_{\text{MUSIC}} : m_{p}+W_{\text{MUSIC}}]$ centered around the peak index $m_p$.
\begin{itemize}
    \item Range Estimation: The spectral covariance matrix is $\mat{R}_R = \frac{1}{N_s} \mat{Y}_{\text{sub}}^T \mat{Y}_{\text{sub}}^*$. Eigen-decomposition yields noise subspace $\mat{U}_{n,R}$. The range estimate is:
        \begin{equation}\label{eq:music_range}
             \hat{r}_k 
             = \argmax_{r} \frac{1}{\|\mat{U}_{n,R}^H \vect{k}_R(r)\|_2^2},
        \end{equation}
         where $[\vect{k}_R(r)]_{w} = \exp(-\j 4\pi r w \Delta f / c)$ and $w$ corresponds to the subcarrier indices within the window used for $\mat{Y}_{\text{sub}}$.
    \item Velocity Estimation: The temporal covariance matrix is $\mat{R}_D = \frac{1}{2W_{\text{MUSIC}}+1} \mat{Y}_{\text{sub}} \mat{Y}_{\text{sub}}^H$. Eigen-decomposition yields the noise subspace $\mat{U}_{n,D}$. The velocity estimate is:
        \begin{equation}\label{eq:music_vel}
            \hat{v}_{r,k} 
            = \argmax_{v} \frac{1}{\|\mat{U}_{n,D}^H \vect{k}_D(v)\|_2^2},
        \end{equation}
        where $[\vect{k}_D(v)]_{n_s} = \exp(\j 4\pi f_{m_p} v n_s T_s / c)$. 
\end{itemize}



\section{Simulation Results}
\label{sec:Simulations}

\subsection{Simulation Setup and CFARNet Training}
\label{subsec:SimSetup}
\begin{table}[t]
\caption{Simulation Parameters}
\label{tab:sim_params}
\centering
\begin{tabularx}{\linewidth}{@{} >{\raggedright\arraybackslash}X l >{\raggedright\arraybackslash}X @{}}
\toprule
\textbf{Parameter} & \textbf{Symbol} & \textbf{Value} \\
\midrule
Start Frequency & $f_0$ & $77$ GHz \\
Center Frequency & $f_c$ & $f_0 + W/2 = 77.5$ GHz \\
Bandwidth & $W$ & $1$ GHz \\
Number of Subcarriers & $M$ & $2048$ \\
Subcarrier Spacing & $\Delta f$ & $W/M \approx 488$ kHz \\
Number of Antenna Elements & $N$ & $128$ \\
Antenna Spacing & $d$ & $\lambda_c/2 = c/2f_c \approx 1.94$ mm \\
Number of OFDM Symbols & $N_s$ & $32$ \\
Angle Scan Start & $\phi_{\text{start}}$ & $-60\degree$ \\
Angle Scan End & $\phi_{\text{end}}$ & $60\degree$ \\
Number of Targets & $K$ & $\{1,2,3,4,5\}$ \\
Target RCS & $\sigma_{c,k}$ & $0.1 \, \text{m}^2$  \\
Tx (Rx) Antenna Gain & $G_T(G_R)$ & $5$ dB  \\
Minimum Angular Separation & $\Delta\phi_{\text{min}}$ & $\{1\degree, 1.5\degree, 3\degree, 5\degree, 10\degree\}$ \\
Transmit Power & $P_t$ & $\{45,50,55,60\}$ dBm \\
MUSIC Sub-band Half-width & $W_{\text{MUSIC}}$ & $10$ \\
\bottomrule
\end{tabularx}
\end{table}

\begin{figure}[t]
    \centering
    \includegraphics[width=1\linewidth]{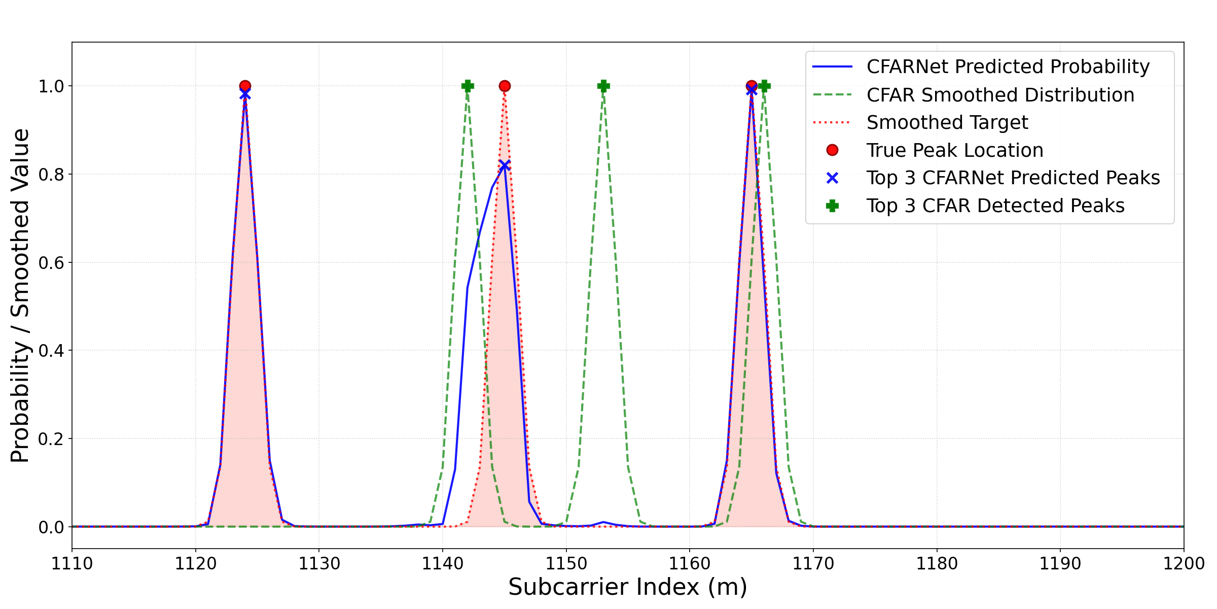} 
    \caption{An example of CNN-based CFARNet output vs. CFAR output vs. Ground truth when there are three close-by targets.}
    \label{fig:prediction_vis}
    \vspace{-3mm}
\end{figure}

The key simulation parameters are summarized in Table~\ref{tab:sim_params}. Datasets were generated with $K$ targets per sample, where each target's parameters were randomly drawn as follows: angle $\phi_k \in [-60^\circ, 60^\circ]$, range $r_k \in [35, 100]$~m, and absolute radial velocity $|v_k| \in [1, 10]$~m/s. All targets were assumed to have identical RCS of $0.1\,\text{m}^2$. A line-of-sight (LoS) channel model was assumed for all targets.

To evaluate the impact of target proximity, separate datasets were generated under different minimum angular separation constraints, specifically $\Delta\phi_{\text{min}} \in \{1^\circ, 1.5^\circ, 3^\circ, 5^\circ, 10^\circ\}$. For each $\Delta\phi_{\text{min}}$, $50,000$ samples were created.
Model training and validation were conducted exclusively on the dataset with $\Delta\phi_{\text{min}} = 1^\circ$. This dataset was partitioned into 35,000 samples for training, 7,500 samples for validation, and 7,500 samples for testing. A single CFARNet model was trained on this data and used for all subsequent evaluations.
Model performance was assessed in two stages: first on the original $\Delta\phi_{\text{min}} = 1^\circ$ test set, and then on independent test sets corresponding to the other $\Delta\phi_{\text{min}}$ values to evaluate generalization capability.

Separate CFARNet models were trained for each transmit power level $P_t \in \{45,50,55,60\}$~dBm. 




\subsection{Performance Evaluation}
\label{subsec:PerfEval}

\subsubsection{CFARNet Output Visualization}
\label{subsubsec:Training} 

Fig.~\ref{fig:prediction_vis} compares CFARNet and traditional CFAR against the ground truth for three targets at $5^{\circ}$, $6^{\circ}$, and $7^{\circ}$, corresponding to subcarrier indices $1124$, $1145$, and $1165$. CFARNet accurately detects all targets, with markers precisely aligning with the ground truth. In contrast, traditional CFAR predicts peaks at $1142$, $1153$, and $1166$, missing the target at $1124$ and producing less accurate estimates around $1145$. This highlights CFARNet’s superior ability to detect and distinguish closely spaced targets where conventional methods may struggle or misidentify targets.



\subsubsection{Comparative Estimation Accuracy}


\begin{figure*}[t!]
    \centering
    \includegraphics[width=\linewidth]{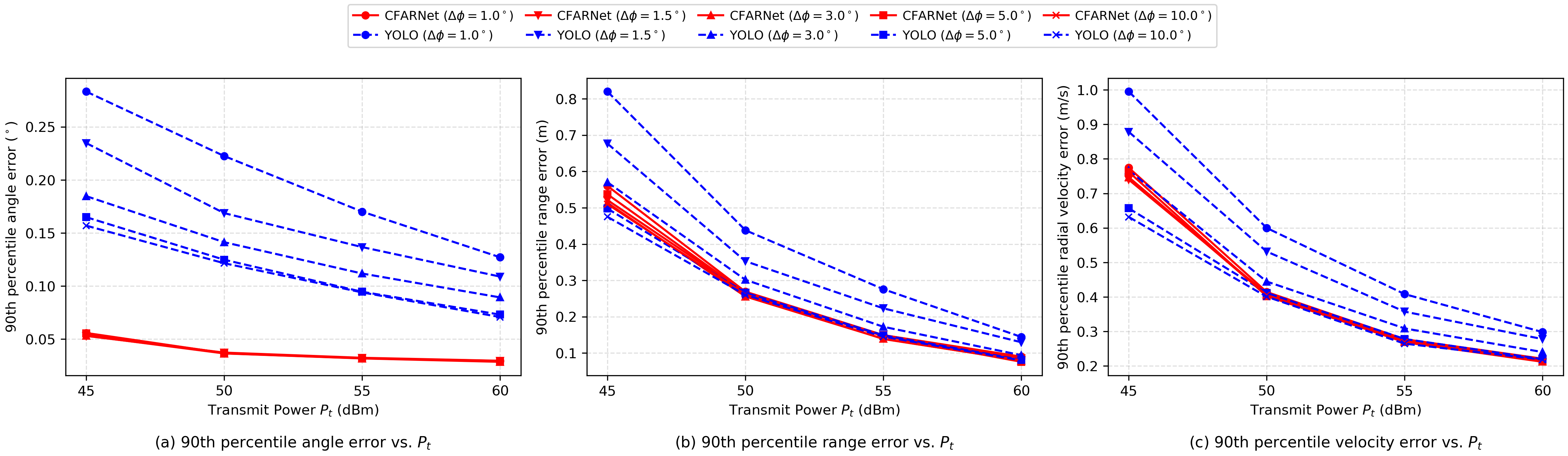}
    
    \caption{90th percentile error performance vs. transmit power for CFARNet and YOLO baseline across different minimum angular separations $\Delta\phi_{\text{min}} \in \{1^\circ, 1.5^\circ, 3^\circ, 5^\circ, 10^\circ\}$.}
    \label{fig:main_results}
    
    \vspace{-3mm} 
\end{figure*}

\begin{figure*}[t!]
    \centering
    \includegraphics[width=\linewidth]{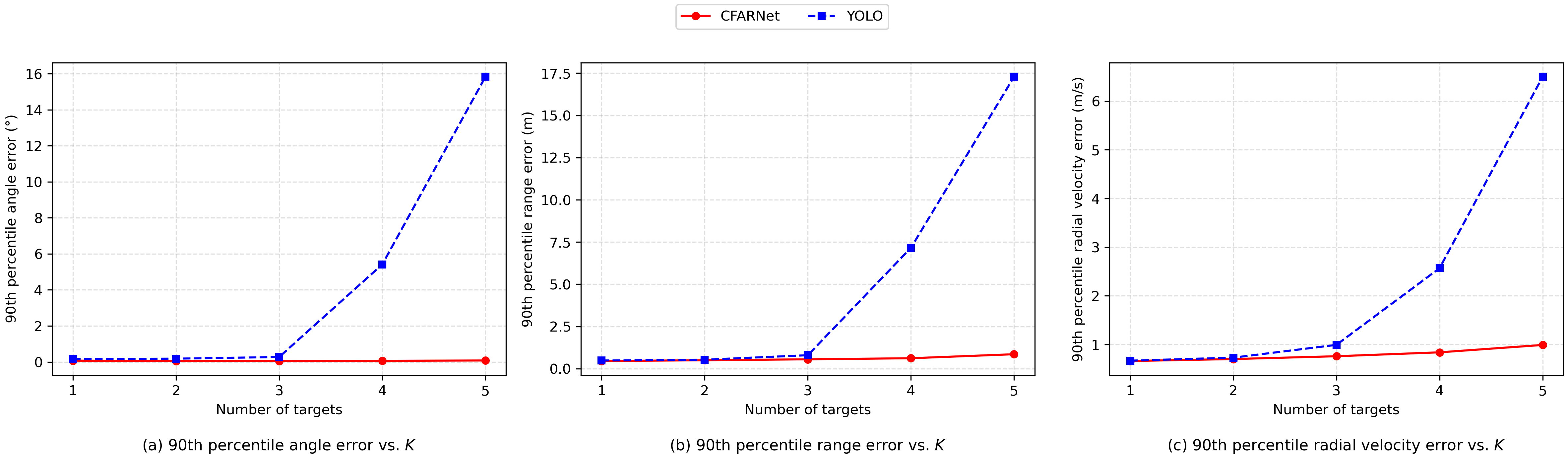}
    \caption{90th percentile error performance vs. number of targets ($K$) in the most challenging scenario where $\Delta\phi_{\min}=1^{\circ}$ and $P_t=45$ dBm.}
    \label{fig:performance_vs_K}
    \vspace{-3mm}
\end{figure*}

Fig.~\ref{fig:main_results} highlights CFARNet's significant advantage in high-resolution scenarios ($\Delta\phi_{\text{min}}=1\degree$) in terms of 90th percentile errors. 
At $P_t=45$~dBm, CFARNet reduces the 90th percentile angle error to $0.056\degree$ compared to $0.283\degree$ for YOLO. 
Similarly, 90th percentile range and velocity errors drop to $0.560$~m and $0.774$~m/s, versus $0.820$~m and $0.995$~m/s. 
This precision advantage persists even at $P_t=60$~dBm (90th percentile angle error $0.029\degree$ vs. $0.127\degree$), demonstrating CFARNet's superior capability in distinguishing closely spaced targets.
For larger angular separations (e.g., $\Delta\phi_{\text{min}} \ge 5\degree$), CFARNet consistently achieves higher angle accuracy. For instance, at $\Delta\phi_{\text{min}} = 5\degree$ and transmit power $P_t = 45$~dBm, CFARNet yields a 90th percentile angle error of $0.056\degree$, significantly lower than YOLO's $0.165\degree$. In contrast, their 90th percentile range and velocity errors are comparable: $0.537$~m vs. $0.498$~m for range, and $0.762$~m/s vs. $0.657$~m/s for velocity, respectively. These results highlight YOLO's competitiveness in range and velocity estimation when targets are well-separated.

Moreover, we investigated the impact of the number of targets in the most challenging scenario of closely spaced targets at low SNR, defined by $\Delta\phi_{\text{min}}=1\degree$ and $P_t = 45$~dBm. Fig.~\ref{fig:performance_vs_K} illustrates that CFARNet consistently outperforms the YOLO baseline regardless of the number of targets. Most notably, as the number of targets increases to five, the YOLO's baseline performance degrades catastrophically: its 90th percentile angle error escalates to $15.83\degree$, its range error reaches $17.28$~m, and its velocity error climbs to $6.50$~m/s. In stark contrast, CFARNet under these same demanding conditions maintains significantly lower errors. Its angle error remains $0.08\degree$, which is about two orders of magnitude lower than YOLO's error. Similarly, its range error is $0.85$~m, over twenty times lower than YOLO's, and its velocity error is $0.99$~m/s, more than six times lower than YOLO's.

These results strongly indicate CFARNet's superior performance and stability when handling multiple, closely spaced targets, especially in the challenging low-SNR scenarios. While the YOLO baseline struggles with feature extraction and peak localization in such dense conditions, CFARNet effectively maintains high estimation accuracy.


The CFARNet demonstrates a significant speed advantage over traditional methods. On our test platform, featuring a dual AMD EPYC 7Y43 CPU (96 cores, 3.63 GHz) and an NVIDIA L40 GPU, CFARNet requires only $7.6$ ms per sample with GPU acceleration, compared to $1.2612$ s for the CPU-based YOLO baseline. Notably, even on the same CPU, CFARNet achieves $75.8$ ms per sample, highlighting its inherent computational efficiency.



\section{Conclusion}
\label{sec:Conclusion}
In this paper, we proposed \textbf{CFARNet}, a deep learning-based processing pipeline for enhanced multi-target parameter estimation in mmWave rainbow beam OFDM radar systems. By replacing the conventional CFAR detector with a CNN designed for peak localization in the angle-Doppler domain, CFARNet effectively addresses the resolution limitations of traditional radar processing, particularly in scenarios with closely spaced targets.

Extensive simulation results demonstrate that CFARNet consistently outperforms the CFAR-based YOLO baseline across various conditions. Most notably, in dense multi-target environments where the baseline performance degrades catastrophically, CFARNet maintains high precision, offering an improvement in angular resolution of up to two orders of magnitude. Furthermore, CFARNet exhibits superior robustness in low-SNR regimes and achieves significantly faster inference speeds compared to traditional CFAR detection, reinforcing its suitability for real-time ISAC applications.

Future work will focus on optimizing the network architecture for further efficiency and scalability. Additionally, we aim to extend this approach to fully end-to-end learning frameworks for joint detection and parameter estimation.
\bibliographystyle{IEEEtran}
\bibliography{reference} 

@article{Gonzalez_VTM25,
	title = {Six {Integration} {Avenues} for {ISAC} in {6G} and {Beyond}},
	volume = {20},
	issn = {1556-6080},
	doi = {10.1109/MVT.2025.3529403},
	number = {1},
	urldate = {2025-03-30},
	journal = {IEEE Vehicular Technology Magazine},
	author = {González-Prelcic, Nuria and Tagliaferri, Dario and Keskin, Musa Furkan and Wymeersch, Henk and Song, Lingyang},
	month = mar,
	year = {2025},
	pages = {18--39},
}

@ARTICLE{Jiang_Yihang_MCOM25,
  author={Jiang, Yihang and Li, Xiaoyang and Zhu, Guangxu and Li, Hang and Deng, Jing and Han, Kaifeng and Shen, Chao and Shi, Qingjiang and Zhang, Rui},
  journal={IEEE Communications Magazine}, 
  title={Integrated Sensing and Communication for Low Altitude Economy: Opportunities and Challenges}, 
  year={2025},
  volume={},
  number={},
  pages={1-7},
  doi={10.1109/MCOM.001.2400685}}

@INPROCEEDINGS{Zhou_Gui_WCNC24,
  author={Zhou, Gui and Peng, Zhendong and Pan, Cunhua and Schober, Robert},
  booktitle={2024 IEEE Wireless Communications and Networking Conference (WCNC)}, 
  title={Rainbow Beams for Wideband mmWave Radar: Beam Training}, 
  year={2024},
  volume={},
  number={},
  pages={1-6},
  doi={10.1109/WCNC57260.2024.10571276}}

@article{Zhou_Gui_JSAC25,
  author={Zhou, Gui and Garkisch, Moritz and Peng, Zhendong and Pan, Cunhua and Schober, Robert},
  journal={IEEE Journal on Selected Areas in Communications}, 
  title={Radar Rainbow Beams for Wideband mmWave Communication: Beam Training and Tracking}, 
  year={2025},
  volume={43},
  number={4},
  pages={1009-1026},
  doi={10.1109/JSAC.2025.3531538}
}

@article{Liu_Fan_JSAC22,
	title = {Integrated {Sensing} and {Communications}: {Toward} {Dual}-{Functional} {Wireless} {Networks} for {6G} and {Beyond}},
	volume = {40},
	issn = {1558-0008},
	shorttitle = {Integrated {Sensing} and {Communications}},
	doi = {10.1109/JSAC.2022.3156632},
	number = {6},
	urldate = {2025-03-11},
	journal = {IEEE Journal on Selected Areas in Communications},
	author = {Liu, Fan and Cui, Yuanhao and Masouros, Christos and Xu, Jie and Han, Tony Xiao and Eldar, Yonina C. and Buzzi, Stefano},
	month = jun,
	year = {2022},
	pages = {1728--1767},
}

@ARTICLE{Wadaskar_TWC25,
  author={Wadaskar, Aditya and Boljanovic, Veljko and Yan, Han and Cabric, Danijela},
  journal={IEEE Transactions on Wireless Communications}, 
  title={{Fast 3D Beam Training With True-Time-Delay Arrays in Wideband Millimeter-Wave Systems}}, 
  year={2025},
  volume={24},
  number={6},
  pages={5146-5162},
  doi={10.1109/TWC.2025.3546157}}

@INPROCEEDINGS{Mo_Jianhua_Asilomar24,
  author={Mo, Jianhua and AlAmmouri, Ahmad and Dong, Shenggang and Nam, Younghan and Choi, Won-Suk and Xu, Gary and Zhang, Jianzhong Charlie},
  booktitle={2024 58th Asilomar Conference on Signals, Systems, and Computers}, 
  title={Beamforming with Joint Phase and Time Array: System Design, Prototyping and Performance}, 
  year={2024},
  volume={},
  number={},
  pages={875-881},
  doi={10.1109/IEEECONF60004.2024.10942922}}

@ARTICLE{Alammouri_Access22,
  author={Alammouri, Ahmad and Mo, Jianhua and Ratnam, Vishnu V. and Ng, Boon Loong and Heath, Robert W. and Lee, Juho and Zhang, Jianzhong},
  journal={IEEE Access}, 
  title={Extending Uplink Coverage of mmWave and Terahertz Systems Through Joint Phase-Time Arrays}, 
  year={2022},
  volume={10},
  number={},
  pages={88872-88884},
  doi={10.1109/ACCESS.2022.3200334}}

@INPROCEEDINGS{Yildiz_VTC24,
  author={Yildiz, Ozlem and AlAmmouri, Ahmad and Mo, Jianhua and Nam, Younghan and Erkip, Elza and Zhang, Jianzhong Charlie},
  booktitle={2024 IEEE 100th Vehicular Technology Conference (VTC2024-Fall)}, 
  title={{3D} Beamforming Through Joint Phase-Time Arrays}, 
  year={2024},
  volume={},
  number={},
  pages={1-7},
  doi={10.1109/VTC2024-Fall63153.2024.10757955}}

@inproceedings{Cai_Yeyue_WCNC25,
author={Cai, Yeyue and Tao, Meixia and Sun, Shu},
  booktitle={2025 IEEE Wireless Communications and Networking Conference (WCNC)}, 
  title={Frequency-Dependent Beamforming for Hybrid Near-Far Field Communications Through Joint Phase-Time Arrays}, 
  year={2025},
  volume={},
  number={},
  pages={1-6},
  doi={10.1109/WCNC61545.2025.10978711}}

@ARTICLE{Forbes_JSSC23,
  author={Forbes, Travis and Magstadt, Benjamin and Moody, Jesse and Saugen, Justine and Suchanek, Andrew and Nelson, Spencer},
  journal={IEEE Journal of Solid-State Circuits}, 
  title={A 0.2–2 {GHz} Time-Interleaved Multistage Switched-Capacitor Delay Element Achieving 2.55–448.6 ns Programmable Delay Range and 330 ns/mm2 Area Efficiency}, 
  year={2023},
  volume={58},
  number={8},
  pages={2349-2359},
  doi={10.1109/JSSC.2023.3257545}}

@article{rotman2016true,
  title={{True time delay in phased arrays}},
  author={Rotman, Ruth and Tur, Moshe and Yaron, Lior},
  journal={Proceedings of the IEEE},
  volume={104},
  number={3},
  pages={504--518},
  year={2016},
  publisher={IEEE}
}

@article{ratnam2022joint,
  title={{Joint phase-time arrays: A paradigm for frequency-dependent analog beamforming in 6G}},
  author={Ratnam, Vishnu V and Mo, Jianhua and Alammouri, Ahmad and Ng, Boon Loong and Zhang, Jianzhong and Molisch, Andreas F},
  journal={IEEE Access},
  volume={10},
  pages={73364--73377},
  year={2022},
  publisher={IEEE}
}

@INPROCEEDINGS{Yan_Han_Asilomar19,
  author={Yan, Han and Boljanovic, Veljko and Cabric, Danijela},
  booktitle={2019 53rd Asilomar Conference on Signals, Systems, and Computers}, 
  title={Wideband Millimeter-Wave Beam Training with True-Time-Delay Array Architecture}, 
  year={2019},
  volume={},
  number={},
  pages={1447-1452},
  doi={10.1109/IEEECONF44664.2019.9048885}}

@article{zhai2021ss,
  title={{SS-OFDMA: Spatial-spread orthogonal frequency division multiple access for terahertz networks}},
  author={Zhai, Bangzhao and Tang, Aimin and Peng, Chen and Wang, Xudong},
  journal={IEEE Journal on Selected Areas in Communications},
  volume={39},
  number={6},
  pages={1678--1692},
  year={2021},
  publisher={IEEE}
}

@ARTICLE{Cui_Mingyao_TWC23,
  author={Cui, Mingyao and Dai, Linglong and Wang, Zhaocheng and Zhou, Shidong and Ge, Ning},
  journal={IEEE Transactions on Wireless Communications}, 
  title={Near-Field Rainbow: Wideband Beam Training for {XL-MIMO}}, 
  year={2023},
  volume={22},
  number={6},
  pages={3899-3912},
  doi={10.1109/TWC.2022.3222198}}

@article{gao2023integrated,
  title={{Integrated sensing and communications with joint beam-squint and beam-split for mmWave/THz massive MIMO}},
  author={Gao, Feifei and Xu, Liangyuan and Ma, Shaodan},
  journal={IEEE Transactions on Communications},
  volume={71},
  number={5},
  pages={2963--2976},
  year={2023},
  publisher={IEEE}
}

@ARTICLE{MUSIC_Schmidt_Ref,
  author={Schmidt, R. O.},
  journal={IEEE Transactions on Antennas and Propagation},
  title={{Multiple emitter location and signal parameter estimation}},
  year={1986},
  volume={34},
  number={3},
  pages={276-280},
  doi={10.1109/TAP.1986.1143830}
}

@ARTICLE{ESPRIT_Roy_Ref,
  author={Roy, R. and Kailath, T.},
  journal={IEEE Transactions on Acoustics, Speech, and Signal Processing}, 
  title={{ESPRIT}-estimation of signal parameters via rotational invariance techniques}, 
  year={1989},
  volume={37},
  number={7},
  pages={984-995},
  doi={10.1109/29.32276}}

@article{CFAR_Limitations_Ref,
  title={Constant false alarm rate detection method in mixed {Weibull} distribution sea clutter},
  author={Mboungam, Abdel Hamid Mbouombouo and Zhi, Yongfeng and Monguen, Cedric Karel Fonzeu},
  journal={Digital Signal Processing},
  volume={149},
  pages={104494},
  year={2024},
  publisher={Elsevier}
}

@article{CFAR_Overview_Ref,
  title={A review of {CFAR} detection techniques in radar systems},
  author={Farina, Alfonso and Studer, Francisco A},
  year={1986},
  publisher={IET}
}

@article{Luo2024yolo,
  title={{YOLO}: An efficient {Terahertz} band integrated sensing and communications scheme with beam squint},
  author={Luo, Hongliang and Gao, Feifei and Lin, Hai and Ma, Shaodan and Poor, H Vincent},
  journal={IEEE Transactions on Wireless Communications},
  volume={23},
  number={8},
  pages={9389--9403},
  year={2024},
  publisher={IEEE}
}

@misc{cai2025hybrid,
      title={Hybrid Near/Far-Field Frequency-Dependent Beamforming via Joint Phase-Time Arrays}, 
      author={Yeyue Cai and Meixia Tao and Jianhua Mo and Shu Sun},
      year={2025},
      eprint={2501.15207},
      archivePrefix={arXiv},
      primaryClass={cs.IT},
      url={https://arxiv.org/abs/2501.15207}, 
}

@article{Mo2019Beam,
  title = {Beam {{Codebook Design}} for {{5G mmWave Terminals}}},
  author = {Mo, Jianhua and Ng, Boon Loong and Chang, Sanghyun and Huang, Pengda and Kulkarni, Mandar N. and Alammouri, Ahmad and Zhang, Jianzhong Charlie and Lee, Jeongheum and Choi, Won-Joon},
  year = {2019},
  journal = {IEEE Access},
  volume = {7},
  pages = {98387--98404},
  issn = {2169-3536},
  doi = {10.1109/ACCESS.2019.2930224},
  urldate = {2022-10-10},
}

@article{Heng2021Six,
  title = {Six {{Key Challenges}} for {{Beam Management}} in 5.{{5G}} and {{6G Systems}}},
  author = {Heng, Yuqiang and Andrews, Jeffrey G. and Mo, Jianhua and Va, Vutha and Ali, Anum and Ng, Boon Loong and Zhang, Jianzhong Charlie},
  year = {2021},
  month = jul,
  journal = {IEEE Communications Magazine},
  volume = {59},
  number = {7},
  pages = {74--79},
  issn = {0163-6804, 1558-1896},
  doi = {10.1109/MCOM.001.2001184},
  urldate = {2022-11-15},
}

@misc{Nam2025Joint,
      title={Joint Phase Time Array: Opportunities, Challenges and System Design Considerations}, 
      author={Young-Han Nam and Ahmad AlAmmouri and Jianhua Mo and Jianzhong Chalrie Zhang},
      year={2025},
      eprint={2412.01714},
      archivePrefix={arXiv},
      primaryClass={eess.SP},
      url={https://arxiv.org/abs/2412.01714}, 
}

@Article{Lei2025Deep,
  author   = {Lei, Hao and Zhang, Jiayi and Xiao, Huahua and Ng, Derrick Wing Kwan and Ai, Bo},
  journal  = {IEEE Transactions on Wireless Communications},
  title    = {{Deep Learning-Based Near-Field User Localization With Beam Squint in Wideband XL-MIMO Systems}},
  year     = {2025},
  number   = {2},
  pages    = {1568-1583},
  volume   = {24},
  doi      = {10.1109/TWC.2024.3510303},
}

@Article{Zheng2025Near,
  author   = {Zheng, Tianyue and Cui, Mingyao and Wu, Zidong and Dai, Linglong},
  journal  = {IEEE Transactions on Wireless Communications},
  title    = {{Near-Field Wideband Beam Training Based on Distance-Dependent Beam Split}},
  year     = {2025},
  number   = {2},
  pages    = {1278-1292},
  volume   = {24},
  doi      = {10.1109/TWC.2024.3507795},
}

\balance 

\end{document}